\documentclass[aps,10pt,prc,tightenlines,twocolumn,twoside,showpacs,nofootinbib]{revtex4}
\usepackage{graphicx,amsmath,amssymb}
\usepackage{hyperref}
\usepackage{breakurl}

\newcommand{\lsim}{\lesssim}
\newcommand{\gsim}{\gtrsim}

\begin{document}
\title{Sequential Regeneration of Charmonia in Heavy-Ion Collisions}

\author{Xiaojian Du\footnote{Corresponding auther.\\E-mail address:xjdu@physics.tamu.edu}} \author{Ralf Rapp} \affiliation{Cyclotron Institute 
  and Department of Physics and Astronomy, Texas A\&M University, College Station, TX
  77843-3366, USA}
\date{\today}

\begin{abstract}
We investigate the production of $\psi(2S)$ in nuclear collisions at RHIC and LHC energies.
We first address charmonium production in 200\,GeV d-Au collisions at RHIC; the strong suppression of 
$\psi'$ mesons observed in these reactions indicates mechanisms beyond initial cold nuclear matter 
effects. We find that a more complete treatment of hadronic dissociation reactions leads to appreciable 
$\psi'$ suppression in the hadronic medium of an expanding fireball background for d-Au collisions. 
When implementing the updated hadronic reaction rates into a fireball for 2.76\,TeV Pb-Pb collisions 
at LHC, the regeneration of $\psi'$ mesons occurs significantly later than for $J/\psi$'s. Despite a 
smaller total number of regenerated $\psi'$, the stronger radial flow at their time of production 
induces a marked enhancement of their $R_{\rm AA}$ relative to $J/\psi$'s in a transverse-momentum range of 
$p_t\simeq$\,3-6\,GeV. We explore the consequences and uncertainties of this ``sequential regeneration"  
mechanism on the $R_{\rm AA}$ double ratio and find that it can reproduce the trends observed in recent CMS 
data. 
\end{abstract}

\maketitle

%%%%%%%%%%%%%%%%%%%%%%%%%%%%%%%%%%%%%%%%%%%%%%%%%%%%%%%%
\section{Introduction}
\label{sec_intro}
%%%%%%%%%%%%%%%%%%%%%%%%%%%%%%%%%%%%%%%%%%%%%%%%%%%%%%%
Charmonium production in ultra-relativistic heavy-ion collisions (URHICs) has been studied 
for over 30 years. The originally proposed $J/\psi$ suppression signature of Quark-Gluon 
Plasma (QGP) formation~\cite{Matsui:1986dk} has evolved into a more complex problem where 
both suppression and so-called regeneration (or statistical hadronization) mechanisms need 
to be considered. Their interplay and relevance depend on collision energy, system size and 
the 3-momentum of the measured charmonia, see, e.g.,
Refs.~\cite{Rapp:2008tf,Kluberg:2009wc,BraunMunzinger:2009ih} for recent reviews. 
The phenomenological modeling of these mechanisms, and their relation to the underlying 
in-medium properties, has progressed significantly in recent years. In particular, kinetic 
transport approaches, when calibrated with existing data from SPS and RHIC, have predicted 
the main features of the $J/\psi$ production observed in the new energy regime at the 
LHC~\cite{Zhao:2011cv,Zhou:2014kka,Song:2011xi} (although significant 
uncertainties due to, e.g., the open-charm cross section persist~\cite{Andronic:2010dt}). 
These include the overall increase of the nuclear modification factor, $R_{\rm AA}$,  
compared to RHIC energies and its enhancement at low transverse momentum, $p_t$.~\cite{Zhao:2007hh}

%~\cite{Gazdzicki:1999rk,Gorenstein:2000ck,Grandchamp:2002iy,Grandchamp:2002wp,Grandchamp:2003uw,BraunMunzinger:1995bp,BraunMunzinger:2001ip,Becattini:2000jw}. 
%Basic charmonium spectral properties from in-medium potential model~\cite{Riek:2010fk} is 
%implemented into a Boltzmann tranport equation as we did in variety of previous work
%[For a most summarized and completed quantative work see~\cite{Zhao:2010nk}]. 

Much less is known about the $2S$ excited state, $\psi'(3686)$. Its small ``binding" energy of
about 45\,MeV (relative to the $D\bar D$ threshold) renders controlled theoretical calculations
of its in-medium properties (binding energy and inelastic reaction rates) challenging.  
Experimentally, the $\psi'$ over $J/\psi$ ratio has been measured at the SPS~\cite{Abreu:1998vw},
where it was found to drop by up to a factor of 3 in central 17.3\,GeV Pb-Pb collisions.
This is consistent with the statistical hadronization approach~\cite{BraunMunzinger:2000px}, 
but it can also be explained by transport approaches with large inelastic reaction rates of 
the $\psi'$ in the hadronic phase~\cite{Sorge:1997bg,Grandchamp:2002wp}. More recently, $\psi'$ 
data have become available for 0.2\,TeV d-Au collisions at RHIC~\cite{Adare:2013ezl} and 
5.02\,TeV p-Pb collisions at LHC~\cite{Abelev:2014zpa}. $\psi'$ mesons were found to be 
significantly more suppressed than $J/\psi$ mesons, which is diffcult to reconcile with initial 
cold-nuclear-matter (CNM) effects since the passing time of the highly Lorentz-contracted 
incoming nuclei is much smaller than the formation time scale of the charmonia. 
Consequently, final-state effects have been put forward to explain these data, e.g., using 
the comover suppression model~\cite{Ferreiro:2014bia}. The latter achieves a good description 
of the collision energy and rapidity dependence of $\psi'$ and $J/\psi$ production 
in d-Au and p-Pb collisions including expected shadowing effects on the parton distribution 
functions (see also Ref.~\cite{Liu:2013via}). 

However, rather unexpected results have emerged from recent measurements by the CMS 
collaboration~\cite{Khachatryan:2014bva} for the double-ratio of the nuclear modification
factor, $R_{\rm AA}$, of $\psi'$ over $J/\psi$ in 2.76\,TeV Pb-Pb collisions at the LHC
(preliminary results are also available from ALICE~\cite{Arnaldi:2012bg}). 
At slightly forward rapidities, 1.6$<$$|y|$$<$2.4, and for transverse momenta 
3$<$$p_t$$<$30\,GeV, this double ratio is around 0.9$\pm$0.45$\pm$0.3 for semi-central 
collisions (similar for peripheral ones), but significantly exceeds one for central 
collisions, 2.3$\pm$0.5$\pm$0.35. Especially the latter has evaded any model 
explanations thus far, see, e.g., the detailed studies in Ref.~\cite{Chen:2013wmr}. 
On the other hand, around midrapidity, and for momenta 6.5$<$$p_t$$<$30\,GeV, a double
ratio of around $\sim$0.5 is found, which is much more in line with common expectations 
of a stronger suppression of $\psi'$ due to its much weaker binding relative to the $J/\psi$.   

In the present paper we put forward a potential mechanism to (partially) resolve the 
above ``puzzle". 
Based on the rather large inelastic reaction rates for the $\psi'$ in hadronic matter
that we deduce from its suppression in d-Au (also in line with the aforementioned 
SPS data), we argue that the inverse reactions of $\psi'$ formation in Pb-Pb collisions
must also happen in the later, hadronic stages of the fireball evolution. In particular,
the $\psi'$ regeneration processes happen later than those for the $J/\psi$ whose
much larger binding energy leads to an earlier ``freezeout" than for the $\psi'$. 
A consequence of such a ``sequential freezeout" is that the collective expansion 
velocity of the medium leads to harder $p_t$ spectra for the $\psi'$. Thus, in terms 
of the $R_{\rm AA}$, the $\psi'$ can outshine the $J/\psi$ in a momentum range of $p_t\gsim M_\psi$, 
which happens to coincide with the lower CMS $p_t$ cut. On the other hand, at higher $p_t$,
the regeneration contribution ceases giving way to a more sequential-like suppression
pattern of primordially produced charmonia.     

Our paper is organized as follow: In Sec.~\ref{sec_hrg} we revisit our hadronic reaction rates 
for both $\psi'$ and $J/\psi$, apply them within a schematic fireball for d-Au collsions at RHIC 
and compare our results to PHENIX data. In Sec.~\ref{sec_seq} we implement the updated hadronic 
rates in our earlier constructed transport approach. In particular, we estimate the time 
windows for sequential charmonium regeneration, and elaborate the uncertainty for the underlying 
charmonium $p_t$ spectra in the context of the CMS data for the $\psi'/J/\psi$ $R_{\rm AA}$ double 
ratio.  We conclude in Sec.~\ref{sec_concl}.

%%%%%%%%%%%%%%%%%%%%%%%%%%%%%%%%%%%%%%%%%%%%%%%%%%%%%%%%
\section{Hadronic Dissociation of Charmonia}
\label{sec_hrg}
%%%%%%%%%%%%%%%%%%%%%%%%%%%%%%%%%%%%%%%%%%%%%%%%%%%%%%%
Dissociation rates of charmonia in hadronic matter are usually considered to be much smaller than 
in the QGP (see, e.g., the discussion in Ref.~\cite{Rapp:2008tf}). However, for the $\psi'$ this 
is not so obvious, since the proximity of its mass to the $D\bar D$ threshold provides a 
large phase space for break-up reactions. 
In the following, we revisit hadronic reactions rates for $J/\psi$ and $\psi'$ mesons
based on effective meson Lagrangians (Sec.~\ref{ssec_rates}) and evaluate their consequences 
for final-state effects in d-Au reactions at RHIC (Sec.~\ref{ssec_pA}).

%%%%%%%%%%%%%%%%%%%%%%%%%%%%%%%%%%%%%%%%%%%%%%%%%%%%%%%%
\subsection{Update on Dissociation Rates}
\label{ssec_rates}
%%%%%%%%%%%%%%%%%%%%%%%%%%%%%%%%%%%%%%%%%%%%%%%%%%%%%%%
Our starting point is the previously employed hadronic reaction rate~\cite{Grandchamp:2002wp} 
based on flavor-$SU(4)$ meson exchange models~\cite{Lin:1999ad,Haglin:2000ar} for the processes 
$J/\psi+\rho\rightarrow D+\bar{D},D^\star+\bar{D}^\star$ (exothermic for 
$m_{J/\psi}+m_\rho > m_D+m_{\bar D}$ and endothermic for 
$m_{J/\psi}+m_\rho < m_{D^\star}+m_{\bar D^\star}$) 
and $J/\psi+\pi\rightarrow D^\star+\bar{D},D+\bar{D}^\star$ (endothermic). For those reactions, 
the $J/\psi$ dissociation rate at $T$=170\,MeV amounts to 1-2\,MeV, corresponding to a lifetime of 
100-200\,fm/$c$.  Even with the typical uncertainties (factor $\sim$2-3) associated with the hadronic 
formfactor cutoff values, this is too small to affect the $J/\psi$ abundance during the 5-10\,fm/c 
lifetime of the hadronic phase in URHICs. For the $\psi'$, geometric scaling by the vacuum radius 
has been assumed, increasing its rates by a factor $(r_{\psi'}/r_{J\psi})^2$, which is approximately 
compatible with constitutent-quark model calculations~\cite{Barnes:2003dg}.

\begin{figure}[!t]
        \centering
        \includegraphics[width=0.48\textwidth]{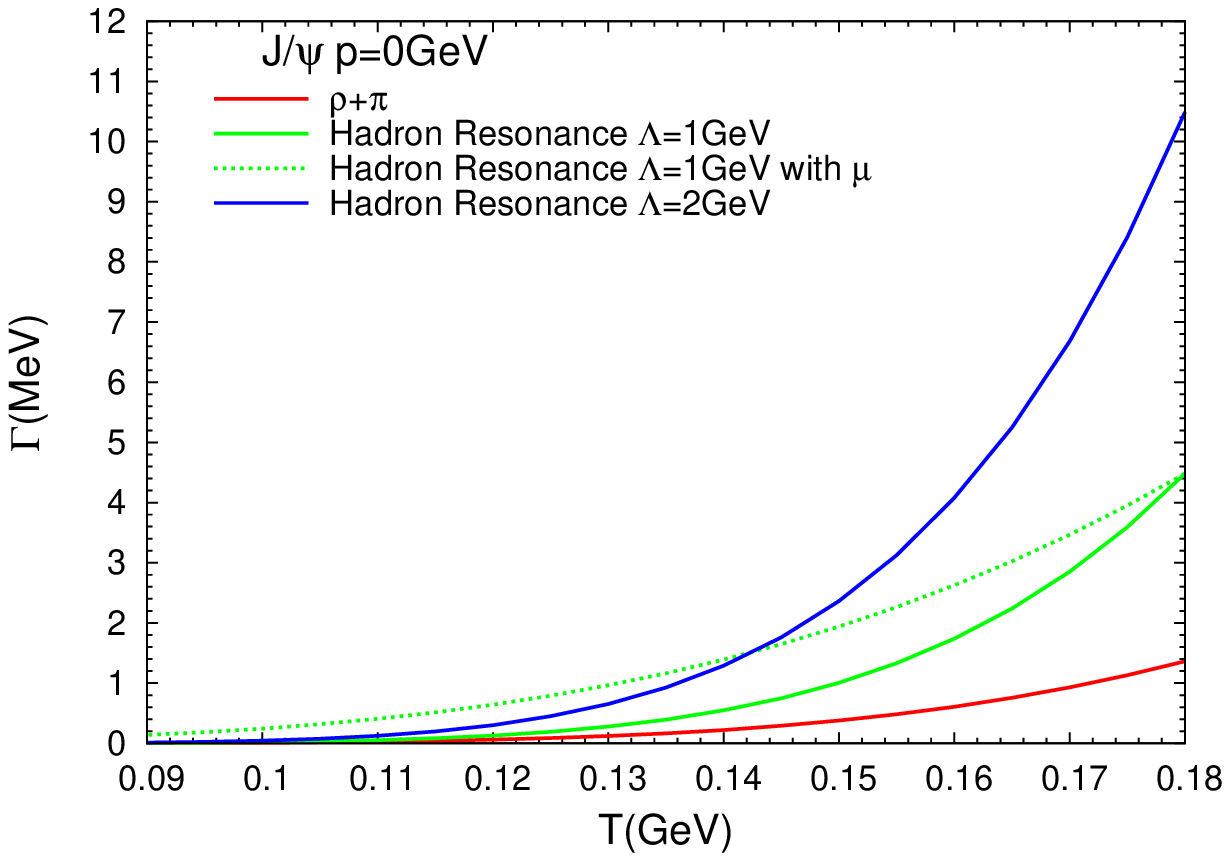}
        \includegraphics[width=0.48\textwidth]{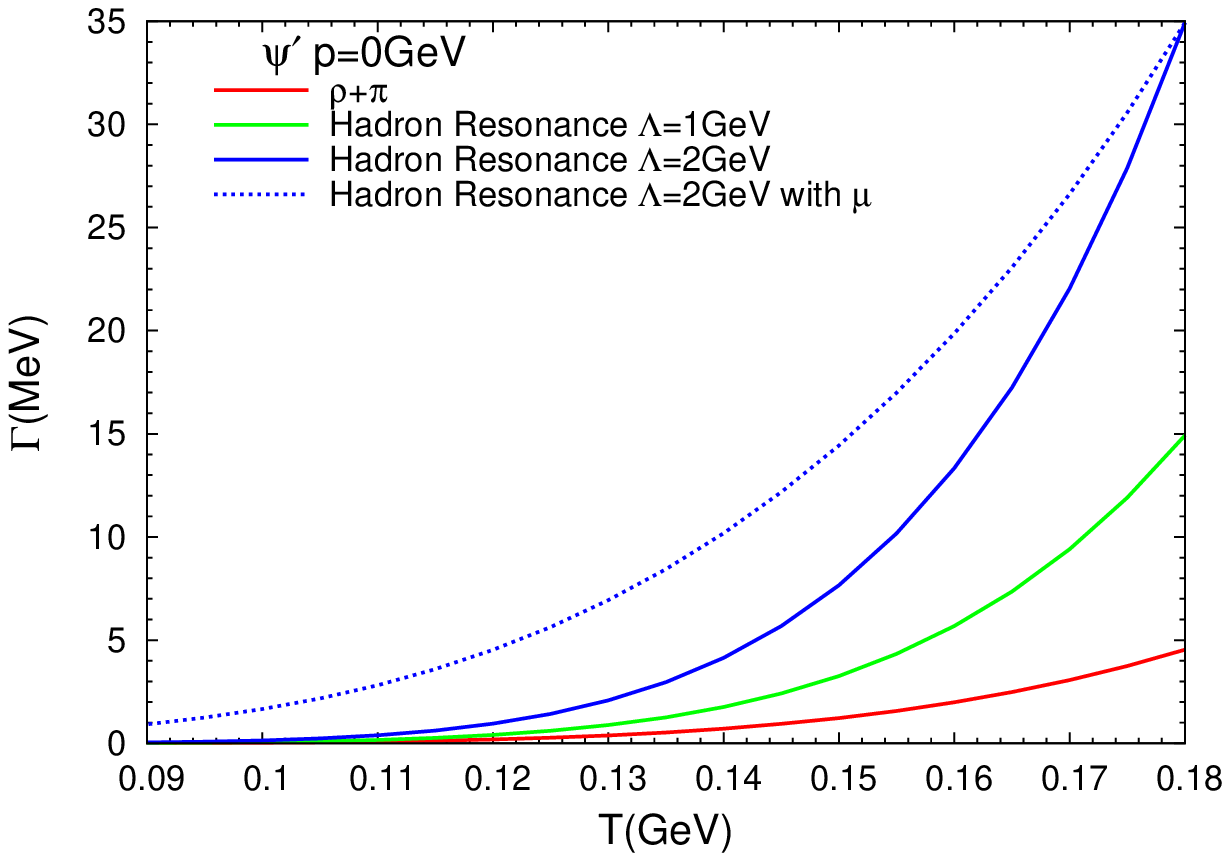}
        \caption{(Color online) Temperature dependence of hadronic dissociation rates
for $J/\psi$ (upper panel) and $\psi'$ (lower panel) at rest in a thermal bath.
Previous results for a $\pi\rho$ gas with $\Lambda$=1\,GeV (solid red lines) are compared to
our updated results for a meson resonance gas using $\Lambda$=1\,GeV and 2\,GeV (solid green
and blue lines, respectively). The blue-dotted lines additionally account for finite meson
chemical potential that build up for temperatures below the chemical freezeout
in URHICs~\cite{Rapp:2002fc}.}
        \label{fig_rHRG}
\end{figure}
A hadron resonance gas (HRG), however, contains many more species than $\pi$ and $\rho$.
To estimate their impact on the charmonium dissociation rates, we simply adopt the 
existing $\rho$- and $\pi$-induced matrix elements and shift their kinematics according 
to the pertinent 2-particle threshold, i.e., 
\begin{equation}
\Gamma^{\rm diss}_{X+J/\psi}(T) = \int \frac{d^3k}{(2\pi)^3} f^X(E_X(k);T) 
\sigma^{\rm in}_{X+J/\psi}(s,s_{\rm thr}^X) v_{\rm rel}  
\end{equation}
with $s=(p_{J/\psi}+k)^2$ and $s_{\rm thr}^X=(m_{J/\psi}+m_X)^2$ for exothermic and 
$s_{\rm thr}^X=(2m_D)^2$ for endothermic channels ($s_{\rm thr}^X=(m_D+m_{D_s})^2$
if $X$ contains a strange quark). We include a total of 52 non-strange and single-strange 
meson species, up to a mass of $m_X=$ 2\,GeV. As before, we apply geometric scaling to
obtain the reaction rates for the $\chi_c$ and $\psi'$.
Our results for a meson gas in chemical equilibrium are summarized by the solid lines
in Fig.~\ref{fig_rHRG}. At the highest temperature ($T$=180\,MeV), the additional
resonances enhance the $J/\psi$ dissociation rate by a factor of $\sim$2.5, and another 
factor of $\sim$2.5 when increasing the hadronic formfactor cutoff from $\Lambda$=1\,GeV to 
2\,GeV, reaching a maximal rate of 10.5\,MeV.  
For the $\psi'$, geometric scaling leads a maximun rate of up to 35\,MeV, translating
into a lifetime of $\sim$6\,fm/$c$ which is now comparable to the duration of the hadronic
phase in URHICs. This becomes even more significant if chemcial freezeout is accounted
for, which implies the build-up of meson chemical potentials leading to a slower
decrease of the meson densities as temperature decreases, cf.~dotted lines in 
Fig.~\ref{fig_rHRG}.

%%%%%%%%%%%%%%%%%%%%%%%%%%%%%%%%%%%%%%%%%%%%%%%%%%%%%%%%
\subsection{Charmonium Production in 0.2\,TeV d-Au Collisions}
\label{ssec_pA}
%%%%%%%%%%%%%%%%%%%%%%%%%%%%%%%%%%%%%%%%%%%%%%%%%%%%%%%
We proceed by implementing our updated hadronic reaction rates into the thermal-rate equation 
framework developed in Ref.~\cite{Grandchamp:2003uw}. Another new aspect relative to our previous 
work~\cite{Zhao:2010nk} is that we allow for final-state effects in small collision systems, 
specifically for d-Au at RHIC, where a rather large $\psi'$ suppression has been observed
(while $J/\psi$'s are much less suppressed).

\begin{figure}[!t]
        \centering
        \includegraphics[width=0.48\textwidth]{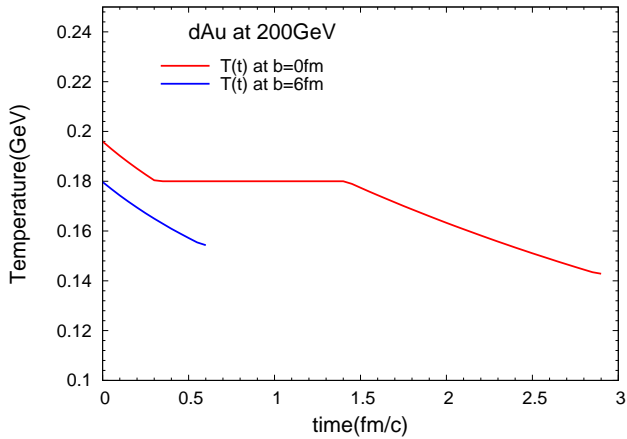}
        \caption{(Color online) Temperature evolution in 0.2\,TeV d-Au collisions assuming a 
thermal fireball model. The blue (red) curve is for an impact parameter of $b$=0(6)\,fm.}
        \label{fig_temp}
\end{figure}
Toward this end, we construct a thermal fireball using the same methods as before for AA 
collisions. We determine the total entropy of the fireball by matching the observed final-state
hadron abundancies. Based on a Glauber model for the centrality dependent nuclear overlap
function, we initialize the fireball with a transverse radius of $R_0$$\simeq$2.5fm/$c$. The 
expansion is then modeled with the same acceleration as in 0.2\,TeV Au-Au before~\cite{Zhao:2010nk}.
For simplicity, we stick with our first-order equation of state, transitioning from a 
quasi-particle QGP into a HRG through a mixed phase at $T$=180\,MeV (we do not expect
large changes when utilizing a modern cross-over EoS, which has been checked for 
dilepton~\cite{Rapp:2013nxa} and bottomonium observables~\cite{Emerick:2011xu}).
Kinetic freezeout is also constructed as before, with a freezeout temperature mildly decreasing 
with centrality, from  $T_{\rm fo}$$\simeq$155\,MeV in peripheral to $\sim$142\,MeV in central 
collisions, resulting in fireball lifetimes of 0.5-3\,fm/$c$, cf.~Fig.~\ref{fig_temp}. For 
the most central collisions, a short QGP phase with initial temperature $T_0$=190\,MeV is 
followed by a 1\,fm/$c$ mixed phase and a 1.5\,fm/$c$ hadronic phase. 

\begin{figure}[!t]
        \centering
        \includegraphics[width=0.48\textwidth]{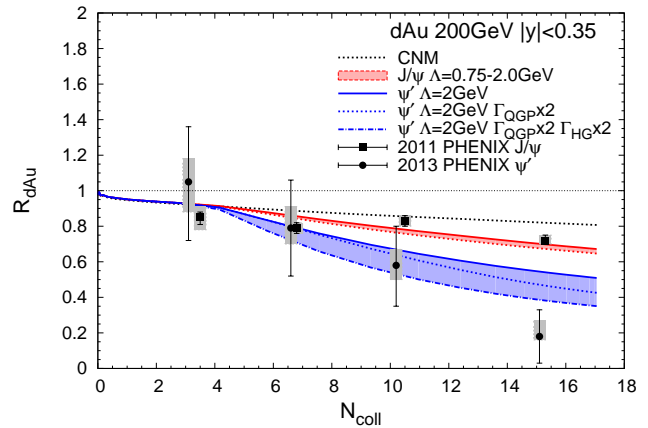}
        \caption{(Color online) Centrality dependence of the nuclear modification factor for
charmonia in 0.2\,TeV d-Au collisions at RHIC; CNM effects for all charmonia are shown by the
black dotted line (consisent with Ref.~\cite{Ferreiro:2014bia} based on the EPS09
shadowing~\cite{Eskola:2009uj} parameterization); the blue band illustrates uncertainties
in both QGP and hadronic dissociation rates of the $\psi'$, while the red band illustrates
the uncertainties due to the hadronic dissociation of the $J/\psi$ (solid (dotted) line: 
$\Lambda$=0.75(2)\,GeV). The PHENIX data are from Refs.~\cite{Adare:2010fn,Adare:2013ezl}.}
        \label{fig_dAuRAA}
\end{figure}
The initial CNM effects are assumed to be identical for all charmonia, essentially given by
a shadowing suppression~\cite{Ferreiro:2014bia} which we mimic with an ``effective" nuclear 
absorption cross section of $\sigma_{\rm abs}$=2.4\,mb. Our results for the centrality 
dependence of the nuclear modification factor,
\begin{equation}
R_{\rm AA}^\Psi(N_{\rm part}) = \frac{N_\Psi(N_{\rm part})}{N_\Psi^{pp} N_{\rm coll}(N_{\rm part})}   \ ,  
\end{equation}
are summarized for both $\Psi=J/\psi, \psi'$ in Fig.~\ref{fig_dAuRAA}, in comparison to PHENIX 
data~\cite{Adare:2013ezl} ($N_{\rm coll}$: number of primordial $NN$ collisions). For the $J/\psi$, 
there is a moderate suppression beyond CNM effects; the additional final-state effects are due to a 
small QGP suppression on the direct $J/\psi$'s, as well as a suppression of the feeddown from 
$\chi_c$'s and $\psi'$, with little room for additional hadronic suppression (thus, in principle, 
preferring a small value for the hadronic cutoff parameter). On the other hand, for the
$\psi'$, our baseline QGP+HRG suppression is not enough to account for the marked suppression
beyond CNM effects. Here, a large formfactor cutoff of $\Lambda$=2\,GeV is preferred to 
augment the hadronic suppression. An additional increase of the QGP suppression rate of the
$\psi'$ by a factor of 2 could also be helpful (such an increase may arise, e.g., from 
nonperturbative heavy-quark interactions with light partons).
%it also renders an approximate scaling of the dissociation rate with entropy density across the
%mixed phase 
Further increasing the hadronic rate by a factor of 2\footnote{This could be due to inelastic 
reactions with baryons and antibaryons or direct $\psi`\to D\bar D$ decays with an in-medium 
reduced $D$ meson mass~\cite{Grandchamp:2003uw}, neither of which we have calculated here.} 
has a similar effect, reducing the discrepancy with the most central PHENIX datum. 

We finally note that a hadronic $\psi'$ dissociation rate well beyond the one 
from the $\pi\rho$ gas (by a factor of 5 or more) was previously deduced within our 
setup~\cite{Grandchamp:2002wp} to be able to account for the $\psi'$ suppression 
observed in S-U and Pb-Pb collisions at the SPS~\cite{Abreu:1998vw}. Our newly
calculated rates in the present ms. are in line with this notion.

%%%%%%%%%%%%%%%%%%%%%%%%%%%%%%%%%%%%%%%%%%%%%%%%%%%%%%%%
\section{Sequential Regeneration of Charmonia}
\label{sec_seq}
%%%%%%%%%%%%%%%%%%%%%%%%%%%%%%%%%%%%%%%%%%%%%%%%%%%%%%%
In this section, we investigate the consequences of the updated hadronic reaction rates 
within our previously constructed thermal fireball expansion (Sec.~\ref{ssec_fireball}),
followed by a more generic evaluation of the associated uncertainties specifically in 
the context of the $\psi'$/$J/\psi$ $R_{\rm AA}$ double ratio (Sec.~\ref{ssec_schematic}). 
We focus on 2.76\,TeV Pb-Pb collisions at the LHC. Our earlier predictions for inclusive 
$J/\psi$ production in these reactions~\cite{Zhao:2011cv} resulted in a fair agreement with 
the centrality, transverse-momentum and rapidiy dependencies observed by the ALICE and CMS 
collaborations, and thus serves as our framework to evaluate $\psi'$ observables. Since 
the pertinent CMS data are for ``prompt" $J/\psi$ and $\psi'$ production, we do not include 
contributions from $B$ feeddown.    

%%%%%%%%%%%%%%%%%%%%%%%%%%%%%%%%%%%%%%%%%%%%%%%%%%%%%%%%
\subsection{Fireball Model}
\label{ssec_fireball}
%%%%%%%%%%%%%%%%%%%%%%%%%%%%%%%%%%%%%%%%%%%%%%%%%%%%%%%
\begin{figure}[!t]
        \centering
        \includegraphics[width=0.48\textwidth]{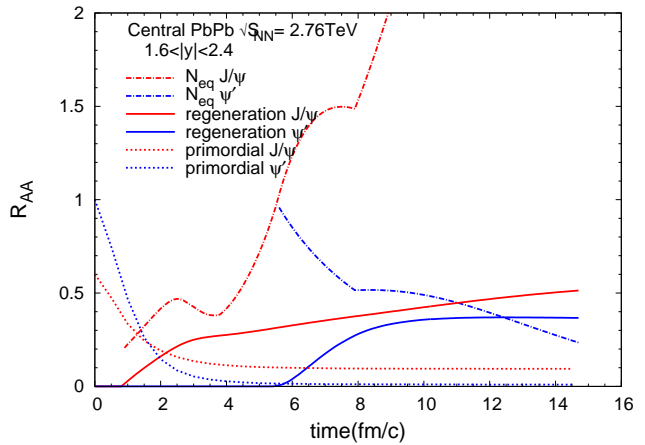}
       \caption{(Color online) Time dependence of $J/\psi$ and $\psi$' nuclear modification factors 
in central($N_{\rm part}$=324) Pb-Pb collisions at $\sqrt{s}$=2.76\,TeV (the denominator for the $J/\psi$ 
$R_{\rm AA}$ includes 40\% feeddown from excited states while the numerator does not). The red (blue) 
curves are for direct $J/\psi$ ($\psi'$), with the solid (dotted) line styles representing the regeneration 
(primordial) contributions; the dash-dotted curves indicate the pertinent equilibrium limits (including
a thermal relaxation time correction~\cite{Grandchamp:2002wp}), starting from the time when the fireball 
has cooled down to the dissociation temperature below which regeneration commences.} 
        \label{fig_time-evo}
\end{figure}

For our fireball results we focus on the so-called ``strong-binding scenario", where
the in-medium charmonium properties are taken with guidance from a $T$-matrix approach~\cite{Riek:2010fk} with
the internal energy from lattice-QCD as underlying potential. This assumption gives a better 
agreement than using the free energy both with correlators from lattice-QCD and the overall
charmonium phenomenology at SPS and RHIC~\cite{Zhao:2010nk}. For definiteness, we  
employ the hadronic rates with $\Lambda$=0.75(2)\,GeV for the $J/\psi$ ($\psi'$), and
a factor of 2 increased QGP rate of the $\psi'$.   

Let us first inspect the time evolution of direct $J/\psi$ and $\psi'$ mesons in 0-20\% 
central Pb-Pb at midrapidity (without shadowing), as following from the solution of the
kinetic rate equation, 
\begin{equation}
\frac{N_\Psi}{d\tau} = - \Gamma_\Psi [N_\Psi (\tau) - N_\psi^{\rm eq} (T(\tau))] \ ,  
\end{equation}
see Fig.~\ref{fig_time-evo}. Compared to our previous results (cf.~lower panel in Fig.~2 of 
Ref.~\cite{Zhao:2011cv}), the $J/\psi$ now picks up a regeneration contribution in the
hadronic phase, by about 0.15 units in $R_{\rm AA}$, which, despite a small rate, is due
to the large equilibrium limit. Most of the production, however, occurs prior to the onset
of the mixed phase at $\tau\simeq5.5$\,fm/$c$. On the contrary, $\psi'$ production only
starts to set in at that point, leveling off at around $\tau$$\simeq$9-10\,fm/$c$, when the
temperature of the fireball has dropped to about 150-160\,MeV. The main qualitative and 
robust feature here is that the lower dissociation temperature of the $\psi'$, relative
to the $J/\psi$, implies a later production through regeneration in the time evolution 
of the fireball in URHICs.       

\begin{figure}[!t]
 \centering
 \includegraphics[width=0.48\textwidth]{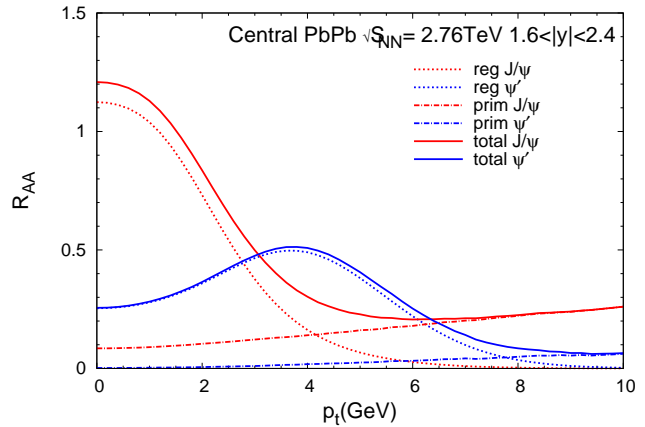}
 \caption{(Color online) Nuclear modification factor for $J/\psi$ (red lines) and $\psi'$ mesons (blue
lines) as a function of $p_t$ for 0-20\% Pb-Pb(2.76\,TeV) collisions. The total $R_{\rm AA}$ for each meson
(solid lines) is decomposed in a regeneration component (dotted lines, evaluated with a blast-wave ansatz in
the fireball evolution at sequential freezeout times) and a suppressed primordial component (dash-dotted
lines, as obtained from a Boltzmann transport equation without gain term), cf.~Ref.~\cite{Zhao:2007hh} for
further details.}
       \label{fig_ptRAA}
\end{figure}

The sequential regeneration of $J/\psi$ and $\psi'$ has rather dramatic consequences on their
transverse-momentum ($p_t$) spectra. Following our previous work~\cite{Zhao:2007hh}, we approximate 
the $p_t$ spectra of the regeneration components with the standard blast-wave expression,
\begin{equation}
\frac{dN^{\rm reg}_\Psi}{p_tdp_t}
= N_0(b) m_t\int_{0}^{R}rdrK_1(\frac{m_t \cosh\rho(r)}{T})I_0(\frac{p_t \sinh\rho(r)}{T})
\label{pt_blastwave}
\end{equation} 
implying thermalized charm-quark distributions. The $m_t=\sqrt{p_t^2+m^2}$ is the transverse mass. As our default, we evaluate this expression 
when most of the pertinent charmonium yield has built up, i.e., at $T_c$ for the $J/\psi$ 
($\tau$=5.5\,fm/$c$ in Fig.~\ref{fig_time-evo}) and at $T$=155\,MeV for the $\psi'$ 
($\tau$=9.7\,fm/$c$ in Fig.~\ref{fig_time-evo}). The resulting nuclear modification factors, $R_{\rm AA}(p_t)$, 
for both regeneration and surviving primordial components are displayed in Fig.~\ref{fig_ptRAA} 
(we have assumed the same initial spectra for $J/\psi$ and $\psi'$ from $pp$ collisions, 
figuring into the denominator of $R_{\rm AA}$). Clearly, there is a significant uncertainty 
associated with this procedure which we address in Sec.~\ref{ssec_schematic} below (we recall, 
however, that our pertinent predictions for the $J/\psi$ gave fair agreement with the observed 
$p_t$ spectra at LHC). 
Three main qualitative features can be gleaned from comparing the $R_{\rm AA}$'s for
$J/\psi$ and $\psi'$ in Fig.~\ref{fig_ptRAA}: 
(a) At low $p_t\lsim 3$\,GeV, the regeneration yield of the $J/\psi$ dominates
over the one from $\psi'$, as a consequence of the approach toward equilibrium which 
favors the smaller $J/\psi$ mass; 
(b) At intermediate $p_t\simeq 3-6$\,GeV, the significantly harder blast wave for 
the $\psi'$  generates a shift of the ``flow bump"  which exceeds the regeneration 
contribution in the $J/\psi$ $R_{\rm AA}$;
(c) at still higher momenta,  $p_t\gsim 6-8$\,GeV, regeneration gives way to (suppressed) 
primordial production, where the $J/\psi$ $R_{\rm AA}$ exceeds again the one from 
the more weakly bound $\psi'$.  
Items (b) and (c) are in qualitative agreement with the trends observed in the 
pertinent $R_{\rm AA}^{\psi'}$/$R_{\rm AA}^{J/\psi}$ double ratio observed by 
CMS~\cite{Khachatryan:2014bva}. In the next section we explore some of the 
uncertainties in our calculations.      

%formation time effect is included. 
%Charged particle multiplicity at forward rapidity $(1.6<y<2.4)$ is extracted from ALICE 
%Collaboration~\cite{Abbas:2013bpa} and the pt-dependence of initi

%%%%%%%%%%%%%%%%%%%%%%%%%%%%%%%%%%%%%%%%%%%%%%%%%%%%%%%%
\subsection{Schematic Model}
\label{ssec_schematic}
%%%%%%%%%%%%%%%%%%%%%%%%%%%%%%%%%%%%%%%%%%%%%%%%%%%%%%%
\begin{figure}[!t]
    \centering
    \includegraphics[width=0.48\textwidth]{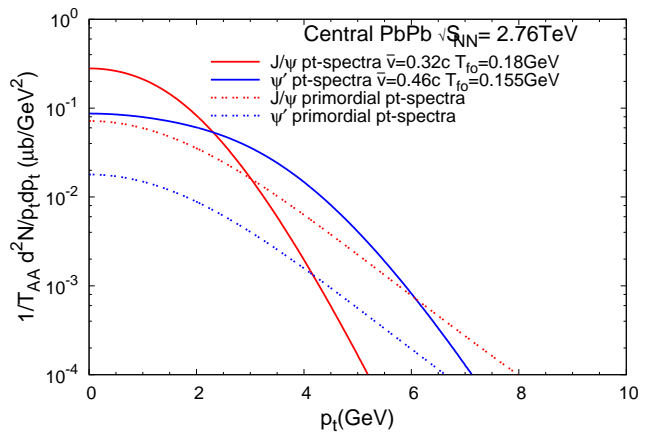}
  \caption{(Color online) Transverse-momentum spectra of $J/\psi$ (red lines) and $\psi'$ mesons 
 (blue lines) for both suppressed primordial (dashed lines, assuming $R_{\rm AA}^{J/\psi}$=0.2 
 and $R_{\rm AA}^{\psi'}$=0.05) and regeneration components (solid lines); the total yields (not 
 the shapes) are normalized to their $N_{\rm coll}$-scaled number in $pp$ collisions.}
        \label{fig_spectra}
\end{figure}
To better quantify variations in the interplay of the different production components of both 
$J/\psi$ and $\psi'$, let us first formulate a baseline scenario motivated by existing 
experimental data for the $J/\psi$ and the thermal fireball calculations in the previous 
sections. For simplicity, we assume the primordial parts to be constant in $R_{\rm AA}(p_t)$. This reflects our currently limited knowledge about the $p_t$ dependence of 
the dissociation rates (see, e.g., Ref.~\cite{Zhao:2007hh}), formation time effects, etc. 
For 0-20\% Pb-Pb collision we take 0.15-0.25 for the $J/\psi$ (compatible with high-$p_t$ CMS 
data for prompt $J/\psi$~\cite{Chatrchyan:2012np}) and 0-0.075 for the $\psi'$, 
%(restricting $R_{\rm AA}^{{\rm prim} \psi'} \le 0.5 R_{\rm AA}^{{\rm prim} J/\psi}$), 
to reflect its stronger absoprtion as a loosely bound state. 
For the regeneration components, we choose total yields such that the total momentum-integrated 
$R_{\rm AA}$ amounts to 0.55-0.65 for the $J/\psi$ (compatible with ALICE data~\cite{Abelev:2013ila}) 
and 0.4 for the $\psi'$ (as suggested by our fireball results). For the sequential freezeout, 
which determines the temperature and flow strength in the blast-wave spectra of the 
regeneration components, we employ the correlation given by the expanding fireball, i.e., 
$\tau\simeq5.5(9.7)$\,fm/$c$ for the $J/\psi$ ($\psi'$) corresponding to $T_{\rm fo}$=180(155)\,MeV 
and $\bar{v}$=0.32(0.46)$c$. Recall that the effective slope parameter for the blast-wave spectra 
is approximately given by $T_{\rm eff}\approx T+m_\Psi\bar{v}^2$, which is mostly driven by the
flow term due to the large mass of the charmonia.
%The average velocity could be evaluated like this $\bar{v}=2a_\bot t_{fo}/3\sqrt{1+(a_\bot t_{fo})^2}$ 
%in the transverse plane.
In Fig.~\ref{fig_spectra} we display the pertinent $p_t$ spectra for central Pb-Pb(2.76\,TeV), 
normalized by the $N_{\rm coll}$-scaled yields in $pp$, to mimic the relative magnitude of $\psi'$ 
and $J/\psi$ contributions in their respective $R_{\rm AA}$'s. The plot highlights again the main effect 
proposed in this paper: due to the stronger flow for the regenerated $\psi'$, its $R_{\rm AA}$ 
can rise above the one for the $J/\psi$ in a limited $p_t$ window around 3-6\,GeV, before
primordial production takes over again. 

\begin{figure}[!t]
	\centering
\includegraphics[width=0.48\textwidth]{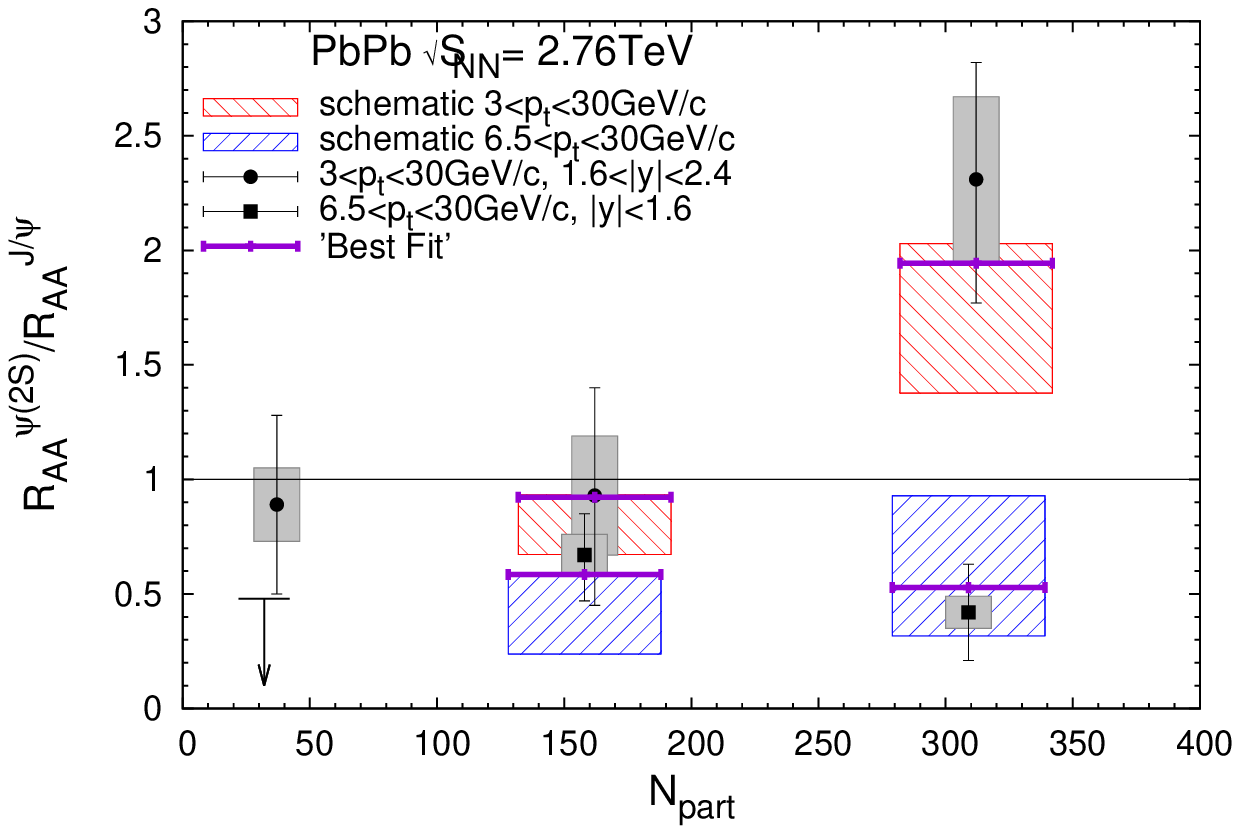}
\includegraphics[width=0.48\textwidth]{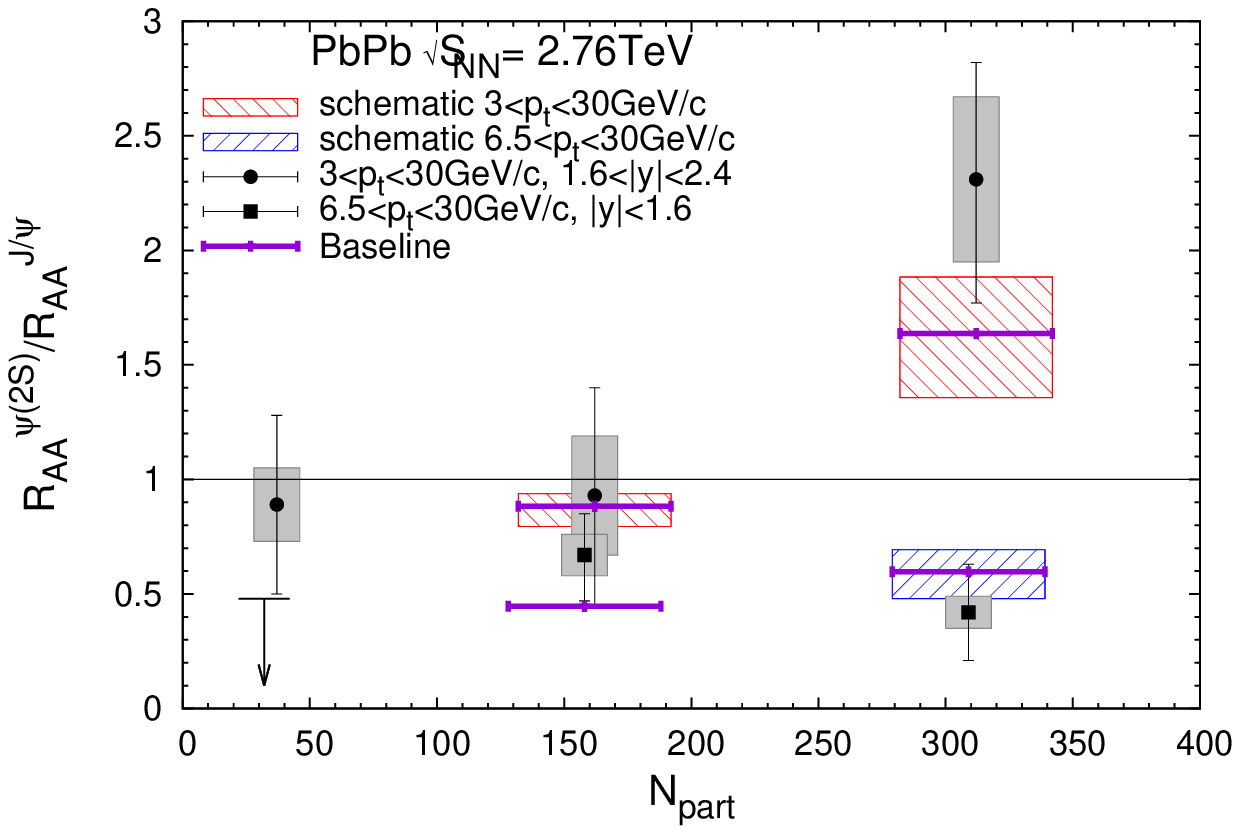}
\caption{(Color online) Ratio of the $R_{\rm AA}$ for $\psi'$ to the one for $J/\psi$ 
(the so-called ``double ratio") as a function of centrality in 2.76\,TeV Pb-Pb collisions. 
The red and blue boxes are our schematic-model results for the double ratio in the momentum 
range $p_t$=3-30\,GeV and $p_t$=6.5-30\,GeV, respectively, and are compared to CMS 
data~\cite{Khachatryan:2014bva}.  
In the upper (lower) panel, the boxes indicate theoretical uncertainties in the modeling
of the primordial (regeneration) component of the charmonium yields and spectra. 
The vertical purple lines represent our ``best fit" in the upper panel, and a ``realistic"
baseline for the blast-wave variations in the lower panel.}
	\label{fig_schem}
\end{figure}

For semi-central Pb-Pb (20-40\%) we construct the baseline following the same reasoning as for
central collisions. We increase the primordial $R_{\rm AA}$'s for $J/\psi$ to 
0.35-0.45~\cite{Chatrchyan:2012np} and to 0.1-0.2 for $\psi'$, and evaluate the blast-wave $p_t$
spectra at the same freezeout temperatures as for central collisions (but with flow velocities 
given by the fireball expansion for 20-40\% centrality).

The resulting double ratios covering the above-specified ranges in the primordial components are
shown in the upper panel of Fig.~\ref{fig_schem} and compared to CMS data~\cite{Khachatryan:2014bva}
(we did not include the ALICE data~\cite{Arnaldi:2012bg} as they contain feeddown contributions 
from $B$-meson decays, which could become rather significant at high $p_t$, see, e.g., 
Ref.~\cite{Chen:2013wmr} for a calculation including those). 
The basic trends of the data can be reproduced within our approach. For defintieness, we quote
(approximate) ``best fit" values of 0.15 and $\sim$0 (0.35 and 0.2) for the primordial $J/\psi$
and $\psi$' values, respectively, in (semi-) central Pb-Pb collisions, resulting in the purple
horizontal bars in the upper panel of Fig.~\ref{fig_schem}. 

Finally, we illustrate the sensitivity of the double ratios to the blast-wave parameters, by varying 
the freezeout temperatures over the ranges $T_{\rm fo}^{J/\psi}$=180-200\,MeV and 
$T_{\rm fo}^{\psi'}$=150-165\,MeV, along with the pertinent flow velocities from the fireball model, 
and fixing the primordial components at $R_{\rm AA}^{J/\psi}$=0.2(0.35) and 
$R_{\rm AA}^{\psi'}$=0.05(0.15) for (semi-) central Pb-Pb. The larger regeneration components and 
larger flow velocities in central collisions render the double ratios more sensitive to 
the details of the sequential regeneration mechanism. 

%%%%%%%%%%%%%%%%%%%%%%%%%%%%%%%%%%%%%%%%%%%%%%%%%%%%%%%%
\section{Conclusions}
\label{sec_concl}
%%%%%%%%%%%%%%%%%%%%%%%%%%%%%%%%%%%%%%%%%%%%%%%%%%%%%%%
In the present work, we have investigated the production systematics of $\psi'$ mesons in 
URHICs. We first revisited the problem of hadronic $\psi'$ dissociation and found that a 
more complete inclusion of hadronic states in a resonance gas suggests a marked increase 
of its inelastic reaction rates. When implementing these rates into an expanding fireball 
for d-Au collisions at RHIC, we found a much improved description of the rather strong 
suppression of $\psi'$ mesons observed in these reactions. This is similar in spirit to, 
and thus supports, the recently suggested comover suppression effects~\cite{Ferreiro:2014bia} 
in dA and $p$A reactions at RHIC and LHC.   
We then evaluated $\psi'$ transport in Pb-Pb collisions at the LHC using our existing fireball 
approach which previously provided fair agreement with $J/\psi$ data. The key
features in our approach are the lower dissociation temperature of the $\psi'$ relative
to the $J/\psi$ and its sizable hadronic reaction rates. This implies a sequential freezeout 
of those two mesons, with most of the $\psi'$ regeneration occurring later in the fireball 
evolution. The larger collective medium flow then leads to an enhancement of the $\psi'$ 
regeneration yield in a $p_t$ region around 3-6\,GeV, transitioning to (suppressed) primordial 
production at higher $p_t$. While quantitative predictions of this mechanism are beyond current 
theoretical control, we have shown that variations in the ingredients to the sequential 
regeneration scenario produce trends in the $\psi'$-over-$J/\psi$ $R_{\rm AA}$ double ratio which 
agree with recent CMS data. We therefore believe that the qualitative features of this mechanism 
are robust and provide a candidate to contribute to the understanding of these data.
In fact, if corroborated, sequential regeneration may serve as a tool to extract in-medium 
properties of the $\psi'$ from URHIC data.

\acknowledgments{We are indebted to Xingbo Zhao for providing us with his codes and valuable discussions.
This work is supported by the US National Science Foundation under grant no. PHY-1306359.} 

\bibliography{ref}

\begin{thebibliography}{33}
\expandafter\ifx\csname natexlab\endcsname\relax\def\natexlab#1{#1}\fi
\expandafter\ifx\csname bibnamefont\endcsname\relax
  \def\bibnamefont#1{#1}\fi
\expandafter\ifx\csname bibfnamefont\endcsname\relax
  \def\bibfnamefont#1{#1}\fi
\expandafter\ifx\csname citenamefont\endcsname\relax
  \def\citenamefont#1{#1}\fi
\expandafter\ifx\csname url\endcsname\relax
  \def\url#1{\texttt{#1}}\fi
\expandafter\ifx\csname urlprefix\endcsname\relax\def\urlprefix{URL }\fi
\providecommand{\bibinfo}[2]{#2}
\providecommand{\eprint}[2][]{\url{#2}}

\bibitem[{\citenamefont{Matsui and Satz}(1986)}]{Matsui:1986dk}
\bibinfo{author}{\bibfnamefont{T.}~\bibnamefont{Matsui}} \bibnamefont{and}
  \bibinfo{author}{\bibfnamefont{H.}~\bibnamefont{Satz}},
  \bibinfo{journal}{Phys. Lett.} \textbf{\bibinfo{volume}{B178}},
  \bibinfo{pages}{416} (\bibinfo{year}{1986}).

\bibitem[{\citenamefont{Rapp et~al.}(2010)\citenamefont{Rapp, Blaschke, and
  Crochet}}]{Rapp:2008tf}
\bibinfo{author}{\bibfnamefont{R.}~\bibnamefont{Rapp}},
  \bibinfo{author}{\bibfnamefont{D.}~\bibnamefont{Blaschke}}, \bibnamefont{and}
  \bibinfo{author}{\bibfnamefont{P.}~\bibnamefont{Crochet}},
  \bibinfo{journal}{Prog. Part. Nucl. Phys.} \textbf{\bibinfo{volume}{65}},
  \bibinfo{pages}{209} (\bibinfo{year}{2010}).

\bibitem[{\citenamefont{Kluberg and Satz}(2010)}]{Kluberg:2009wc}
\bibinfo{author}{\bibfnamefont{L.}~\bibnamefont{Kluberg}} \bibnamefont{and}
  \bibinfo{author}{\bibfnamefont{H.}~\bibnamefont{Satz}},
  \bibinfo{journal}{Landolt-Bornstein} \textbf{\bibinfo{volume}{23}},
  \bibinfo{pages}{372} (\bibinfo{year}{2010}), \eprint{arXiv:0901.3831}.

\bibitem[{\citenamefont{Braun-Munzinger and
  Stachel}(2010)}]{BraunMunzinger:2009ih}
\bibinfo{author}{\bibfnamefont{P.}~\bibnamefont{Braun-Munzinger}}
  \bibnamefont{and} \bibinfo{author}{\bibfnamefont{J.}~\bibnamefont{Stachel}},
  \bibinfo{journal}{Landolt-Bornstein} \textbf{\bibinfo{volume}{23}},
  \bibinfo{pages}{424} (\bibinfo{year}{2010}), \eprint{arXiv:0901.2500}.

\bibitem[{\citenamefont{Zhao and Rapp}(2011)}]{Zhao:2011cv}
\bibinfo{author}{\bibfnamefont{X.}~\bibnamefont{Zhao}} \bibnamefont{and}
  \bibinfo{author}{\bibfnamefont{R.}~\bibnamefont{Rapp}},
  \bibinfo{journal}{Nucl. Phys.} \textbf{\bibinfo{volume}{A859}},
  \bibinfo{pages}{114} (\bibinfo{year}{2011}).

\bibitem[{\citenamefont{Zhou et~al.}(2014)\citenamefont{Zhou, Xu, Xu, and
  Zhuang}}]{Zhou:2014kka}
\bibinfo{author}{\bibfnamefont{K.}~\bibnamefont{Zhou}},
  \bibinfo{author}{\bibfnamefont{N.}~\bibnamefont{Xu}},
  \bibinfo{author}{\bibfnamefont{Z.}~\bibnamefont{Xu}}, \bibnamefont{and}
  \bibinfo{author}{\bibfnamefont{P.}~\bibnamefont{Zhuang}},
  \bibinfo{journal}{Phys. Rev.} \textbf{\bibinfo{volume}{C89}},
  \bibinfo{pages}{054911} (\bibinfo{year}{2014}).

\bibitem[{\citenamefont{Song et~al.}(2011)\citenamefont{Song, Han, and
  Ko}}]{Song:2011xi}
\bibinfo{author}{\bibfnamefont{T.}~\bibnamefont{Song}},
  \bibinfo{author}{\bibfnamefont{K.~C.} \bibnamefont{Han}}, \bibnamefont{and}
  \bibinfo{author}{\bibfnamefont{C.~M.} \bibnamefont{Ko}},
  \bibinfo{journal}{Phys. Rev.} \textbf{\bibinfo{volume}{C84}},
  \bibinfo{pages}{034907} (\bibinfo{year}{2011}).

\bibitem[{\citenamefont{Andronic et~al.}(2010)\citenamefont{Andronic,
  Braun-Munzinger, Redlich, and Stachel}}]{Andronic:2010dt}
\bibinfo{author}{\bibfnamefont{A.}~\bibnamefont{Andronic}},
  \bibinfo{author}{\bibfnamefont{P.}~\bibnamefont{Braun-Munzinger}},
  \bibinfo{author}{\bibfnamefont{K.}~\bibnamefont{Redlich}}, \bibnamefont{and}
  \bibinfo{author}{\bibfnamefont{J.}~\bibnamefont{Stachel}},
  \bibinfo{journal}{J. Phys.} \textbf{\bibinfo{volume}{G37}},
  \bibinfo{pages}{094014} (\bibinfo{year}{2010}).

\bibitem[{\citenamefont{Zhao and Rapp}(2008)}]{Zhao:2007hh}
\bibinfo{author}{\bibfnamefont{X.}~\bibnamefont{Zhao}} \bibnamefont{and}
  \bibinfo{author}{\bibfnamefont{R.}~\bibnamefont{Rapp}},
  \bibinfo{journal}{Phys. Lett.} \textbf{\bibinfo{volume}{B664}},
  \bibinfo{pages}{253} (\bibinfo{year}{2008}).

\bibitem[{\citenamefont{Abreu et~al.}(1998)}]{Abreu:1998vw}
\bibinfo{author}{\bibfnamefont{M.}~\bibnamefont{Abreu}} \bibnamefont{et~al.}
  (\bibinfo{collaboration}{NA50}), \bibinfo{journal}{Nucl. Phys.}
  \textbf{\bibinfo{volume}{A638}}, \bibinfo{pages}{261} (\bibinfo{year}{1998}).

\bibitem[{\citenamefont{Braun-Munzinger and
  Stachel}(2000)}]{BraunMunzinger:2000px}
\bibinfo{author}{\bibfnamefont{P.}~\bibnamefont{Braun-Munzinger}}
  \bibnamefont{and} \bibinfo{author}{\bibfnamefont{J.}~\bibnamefont{Stachel}},
  \bibinfo{journal}{Phys. Lett.} \textbf{\bibinfo{volume}{B490}},
  \bibinfo{pages}{196} (\bibinfo{year}{2000}).

\bibitem[{\citenamefont{Sorge et~al.}(1997)\citenamefont{Sorge, Shuryak, and
  Zahed}}]{Sorge:1997bg}
\bibinfo{author}{\bibfnamefont{H.}~\bibnamefont{Sorge}},
  \bibinfo{author}{\bibfnamefont{E.~V.} \bibnamefont{Shuryak}},
  \bibnamefont{and} \bibinfo{author}{\bibfnamefont{I.}~\bibnamefont{Zahed}},
  \bibinfo{journal}{Phys. Rev. Lett.} \textbf{\bibinfo{volume}{79}},
  \bibinfo{pages}{2775} (\bibinfo{year}{1997}).

\bibitem[{\citenamefont{Grandchamp and Rapp}(2002)}]{Grandchamp:2002wp}
\bibinfo{author}{\bibfnamefont{L.}~\bibnamefont{Grandchamp}} \bibnamefont{and}
  \bibinfo{author}{\bibfnamefont{R.}~\bibnamefont{Rapp}},
  \bibinfo{journal}{Nucl. Phys.} \textbf{\bibinfo{volume}{A709}},
  \bibinfo{pages}{415} (\bibinfo{year}{2002}).

\bibitem[{\citenamefont{Adare et~al.}(2013)}]{Adare:2013ezl}
\bibinfo{author}{\bibfnamefont{A.}~\bibnamefont{Adare}} \bibnamefont{et~al.}
  (\bibinfo{collaboration}{PHENIX}), \bibinfo{journal}{Phys. Rev. Lett.}
  \textbf{\bibinfo{volume}{111}}, \bibinfo{pages}{202301}
  (\bibinfo{year}{2013}).

\bibitem[{\citenamefont{Abelev et~al.}(2014{\natexlab{a}})}]{Abelev:2014zpa}
\bibinfo{author}{\bibfnamefont{B.~B.} \bibnamefont{Abelev}}
  \bibnamefont{et~al.} (\bibinfo{collaboration}{ALICE}),
  \bibinfo{journal}{JHEP} \textbf{\bibinfo{volume}{1412}}, \bibinfo{pages}{073}
  (\bibinfo{year}{2014}{\natexlab{a}}).

\bibitem[{\citenamefont{Ferreiro}(2014)}]{Ferreiro:2014bia}
\bibinfo{author}{\bibfnamefont{E.}~\bibnamefont{Ferreiro}}
  (\bibinfo{year}{2014}), \eprint{arXiv:1411.0549}.

\bibitem[{\citenamefont{Liu et~al.}(2014)\citenamefont{Liu, Ko, and
  Song}}]{Liu:2013via}
\bibinfo{author}{\bibfnamefont{Y.}~\bibnamefont{Liu}},
  \bibinfo{author}{\bibfnamefont{C.~M.} \bibnamefont{Ko}}, \bibnamefont{and}
  \bibinfo{author}{\bibfnamefont{T.}~\bibnamefont{Song}},
  \bibinfo{journal}{Phys. Lett.} \textbf{\bibinfo{volume}{B728}},
  \bibinfo{pages}{437} (\bibinfo{year}{2014}).

\bibitem[{\citenamefont{Khachatryan et~al.}(2014)}]{Khachatryan:2014bva}
\bibinfo{author}{\bibfnamefont{V.}~\bibnamefont{Khachatryan}}
  \bibnamefont{et~al.} (\bibinfo{collaboration}{CMS}), \bibinfo{journal}{Phys.
  Rev. Lett.} \textbf{\bibinfo{volume}{113}}, \bibinfo{pages}{262301}
  (\bibinfo{year}{2014}).

\bibitem[{\citenamefont{Arnaldi}(2013)}]{Arnaldi:2012bg}
\bibinfo{author}{\bibfnamefont{R.}~\bibnamefont{Arnaldi}}
  (\bibinfo{collaboration}{ALICE}), \bibinfo{journal}{Nucl. Phys.}
  \textbf{\bibinfo{volume}{A904-905}}, \bibinfo{pages}{595c}
  (\bibinfo{year}{2013}).

\bibitem[{\citenamefont{Chen et~al.}(2013)\citenamefont{Chen, Liu, Zhou, and
  Zhuang}}]{Chen:2013wmr}
\bibinfo{author}{\bibfnamefont{B.}~\bibnamefont{Chen}},
  \bibinfo{author}{\bibfnamefont{Y.}~\bibnamefont{Liu}},
  \bibinfo{author}{\bibfnamefont{K.}~\bibnamefont{Zhou}}, \bibnamefont{and}
  \bibinfo{author}{\bibfnamefont{P.}~\bibnamefont{Zhuang}},
  \bibinfo{journal}{Phys. Lett.} \textbf{\bibinfo{volume}{B726}},
  \bibinfo{pages}{725} (\bibinfo{year}{2013}).

\bibitem[{\citenamefont{Lin and Ko}(2000)}]{Lin:1999ad}
\bibinfo{author}{\bibfnamefont{Z.-w.} \bibnamefont{Lin}} \bibnamefont{and}
  \bibinfo{author}{\bibfnamefont{C.}~\bibnamefont{Ko}}, \bibinfo{journal}{Phys.
  Rev.} \textbf{\bibinfo{volume}{C62}}, \bibinfo{pages}{034903}
  (\bibinfo{year}{2000}).

\bibitem[{\citenamefont{Haglin and Gale}(2001)}]{Haglin:2000ar}
\bibinfo{author}{\bibfnamefont{K.~L.} \bibnamefont{Haglin}} \bibnamefont{and}
  \bibinfo{author}{\bibfnamefont{C.}~\bibnamefont{Gale}},
  \bibinfo{journal}{Phys. Rev.} \textbf{\bibinfo{volume}{C63}},
  \bibinfo{pages}{065201} (\bibinfo{year}{2001}).

\bibitem[{\citenamefont{Barnes et~al.}(2003)\citenamefont{Barnes, Swanson,
  Wong, and Xu}}]{Barnes:2003dg}
\bibinfo{author}{\bibfnamefont{T.}~\bibnamefont{Barnes}},
  \bibinfo{author}{\bibfnamefont{E.}~\bibnamefont{Swanson}},
  \bibinfo{author}{\bibfnamefont{C.}~\bibnamefont{Wong}}, \bibnamefont{and}
  \bibinfo{author}{\bibfnamefont{X.}~\bibnamefont{Xu}}, \bibinfo{journal}{Phys.
  Rev.} \textbf{\bibinfo{volume}{C68}}, \bibinfo{pages}{014903}
  (\bibinfo{year}{2003}).

\bibitem[{\citenamefont{Rapp}(2002)}]{Rapp:2002fc}
\bibinfo{author}{\bibfnamefont{R.}~\bibnamefont{Rapp}}, \bibinfo{journal}{Phys.
  Rev.} \textbf{\bibinfo{volume}{C66}}, \bibinfo{pages}{017901}
  (\bibinfo{year}{2002}).

\bibitem[{\citenamefont{Grandchamp et~al.}(2004)\citenamefont{Grandchamp, Rapp,
  and Brown}}]{Grandchamp:2003uw}
\bibinfo{author}{\bibfnamefont{L.}~\bibnamefont{Grandchamp}},
  \bibinfo{author}{\bibfnamefont{R.}~\bibnamefont{Rapp}}, \bibnamefont{and}
  \bibinfo{author}{\bibfnamefont{G.~E.} \bibnamefont{Brown}},
  \bibinfo{journal}{Phys. Rev. Lett.} \textbf{\bibinfo{volume}{92}},
  \bibinfo{pages}{212301} (\bibinfo{year}{2004}).

\bibitem[{\citenamefont{Zhao and Rapp}(2010)}]{Zhao:2010nk}
\bibinfo{author}{\bibfnamefont{X.}~\bibnamefont{Zhao}} \bibnamefont{and}
  \bibinfo{author}{\bibfnamefont{R.}~\bibnamefont{Rapp}},
  \bibinfo{journal}{Phys. Rev.} \textbf{\bibinfo{volume}{C82}},
  \bibinfo{pages}{064905} (\bibinfo{year}{2010}).

\bibitem[{\citenamefont{Rapp}(2013)}]{Rapp:2013nxa}
\bibinfo{author}{\bibfnamefont{R.}~\bibnamefont{Rapp}}, \bibinfo{journal}{Adv.
  High Energy Phys.} \textbf{\bibinfo{volume}{2013}}, \bibinfo{pages}{148253}
  (\bibinfo{year}{2013}).

\bibitem[{\citenamefont{Emerick et~al.}(2012)\citenamefont{Emerick, Zhao, and
  Rapp}}]{Emerick:2011xu}
\bibinfo{author}{\bibfnamefont{A.}~\bibnamefont{Emerick}},
  \bibinfo{author}{\bibfnamefont{X.}~\bibnamefont{Zhao}}, \bibnamefont{and}
  \bibinfo{author}{\bibfnamefont{R.}~\bibnamefont{Rapp}},
  \bibinfo{journal}{Eur. Phys. J.} \textbf{\bibinfo{volume}{A48}},
  \bibinfo{pages}{72} (\bibinfo{year}{2012}).

\bibitem[{\citenamefont{Eskola et~al.}(2009)\citenamefont{Eskola, Paukkunen,
  and Salgado}}]{Eskola:2009uj}
\bibinfo{author}{\bibfnamefont{K.}~\bibnamefont{Eskola}},
  \bibinfo{author}{\bibfnamefont{H.}~\bibnamefont{Paukkunen}},
  \bibnamefont{and} \bibinfo{author}{\bibfnamefont{C.}~\bibnamefont{Salgado}},
  \bibinfo{journal}{JHEP} \textbf{\bibinfo{volume}{0904}}, \bibinfo{pages}{065}
  (\bibinfo{year}{2009}).

\bibitem[{\citenamefont{Adare et~al.}(2011)}]{Adare:2010fn}
\bibinfo{author}{\bibfnamefont{A.}~\bibnamefont{Adare}} \bibnamefont{et~al.}
  (\bibinfo{collaboration}{PHENIX}), \bibinfo{journal}{Phys. Rev. Lett.}
  \textbf{\bibinfo{volume}{107}}, \bibinfo{pages}{142301}
  (\bibinfo{year}{2011}).

\bibitem[{\citenamefont{Riek and Rapp}(2010)}]{Riek:2010fk}
\bibinfo{author}{\bibfnamefont{F.}~\bibnamefont{Riek}} \bibnamefont{and}
  \bibinfo{author}{\bibfnamefont{R.}~\bibnamefont{Rapp}},
  \bibinfo{journal}{Phys. Rev.} \textbf{\bibinfo{volume}{C82}},
  \bibinfo{pages}{035201} (\bibinfo{year}{2010}).

\bibitem[{\citenamefont{Chatrchyan et~al.}(2012)}]{Chatrchyan:2012np}
\bibinfo{author}{\bibfnamefont{S.}~\bibnamefont{Chatrchyan}}
  \bibnamefont{et~al.} (\bibinfo{collaboration}{CMS}), \bibinfo{journal}{JHEP}
  \textbf{\bibinfo{volume}{1205}}, \bibinfo{pages}{063} (\bibinfo{year}{2012}).

\bibitem[{\citenamefont{Abelev et~al.}(2014{\natexlab{b}})}]{Abelev:2013ila}
\bibinfo{author}{\bibfnamefont{B.~B.} \bibnamefont{Abelev}}
  \bibnamefont{et~al.} (\bibinfo{collaboration}{ALICE}),
  \bibinfo{journal}{Phys. Lett.} \textbf{\bibinfo{volume}{B734}},
  \bibinfo{pages}{314} (\bibinfo{year}{2014}{\natexlab{b}}).

\end{thebibliography}

\end{document}